\newif\ifspe \spefalse
\ifspe
\documentclass[times]{speauth}
\else
\documentclass[10pt,a4paper]{scrartcl}

\def\HgReadableDate{2015 May 17}
\usepackage{amstext}
\fi

\title{Xoxa: a lightweight approach to normalizing and signing XML}

\ifspe
\author{Norman Gray\affil1\affil2}
\date{XXX 2014}
\else
\author{Norman Gray\thanks{School of Physics and Astronomy, University of
  Glasgow, Glasgow G12 8QQ, UK \url{http://nxg.me.uk}; and Octameter Computing \url{http://8ameter.com}}}
\date{\HgReadableDate}
\fi

\long\def\theabstract{Cryptographically signing XML, and normalizing
it prior to signing,
are forbiddingly intricate problems in the general case.  This is
largely because of the complexities of the XML Information Set.  We
can define a more aggressive normalization, which dispenses with
distinctions and features
which are unimportant in a large class of cases, and thus define a
straightforwardly implementable and portable signature framework.

To be submitted to \textit{Software: Practice and Experience}.

Copyright 2015 Norman Gray.
}

\ifspe
\runningheads{N Gray}{Normalizing and signing XML}
\address{\affilnum1 School of Physics and Astronomy, University of
  Glasgow, Glasgow G12 8QQ, UK \url{http://nxg.me.uk}; \affilnum2~Octameter
  Computing \url{http://8ameter.com}}
\begin{abstract}
\theabstract
\end{abstract}
\fi

\usepackage{url}
\usepackage{hyperref}

\newcommand\tableref[1]{Table~\ref{#1}}
\newcommand\figref[1]{Fig.\,\ref{#1}}
\newcommand\secref[1]{Sect.\,\ref{#1}}

\newcommand\Norm{\mathop{\mathrm{Norm}}}
\newcommand\ESIS{\mathop{\mathrm{ESIS}}}
\newcommand\Sign{\mathop{\mathrm{Sign}}}

\begin{document}
\maketitle

\ifspe\else
\begin{abstract}
\theabstract
\end{abstract}
\fi

\section{Introduction}

Both normalizing and signing XML appear to be hard problems, given the
size and complexity of the work of the W3C Signature working
group,\footnote{\url{http://www.w3.org/Signature/}} which has produced
recommendations on creating signatures for XML, as
well as on the necessary prior problem of canonicalizing
XML.

The goal of such signing algorithms is to provide the assurance that
an XML document read from a file, received in a message, or retrieved from an
XML database, is the same as the document that was written, sent, or
uploaded, and that it has not been changed deliberately or
accidentally.  While reading and writing a file are straightforward,
it can sometimes be hard in practice to ensure that the sequence of bytes
deserialized from a network stack, or re-serialized from a database,
are exactly the same as the original ones, rather than a sequence of
bytes deemed equivalent by the rules of XML.

For each XML document, there is a set of other documents with
trivially different syntax -- that is, they are equivalent for all
practical purposes, but use single-quotes rather than double-quotes
for marking attributes, or have the attributes in a different order,
or have ignorable (according to a DTD) whitespace differences between
elements, or they appear in a different encoding.  The process of
canonicalization consists of precisely defining and operationalizing
this equivalence, and selecting one of the documents in that set as
the canonical representative of it.  By this means an XML document can
be signed by signing the canonically equivalent document; this
signature is then valid for any other member of the equivalence class.

The W3C signature framework is complicated, firstly because it
supports an intricate mechanism for specifying the signing of rather
general transformations of the input document; and secondly because by
aiming to make the equivalence class as small as
possible, it must be sensitive to many of the unavoidably intricate
details of the low-level syntax of XML documents.  The complication
and the sensitivity make XML canonicalization hair-pullingly fragile.

This paper presents an alternative approach.  By choosing to perform
the canonicalization step at a more abstract level, we can make that
step both simpler to specify and simpler to implement, with a
resulting object which may be very straightforwardly signed.  The
resulting set of equivalent documents will be larger than for the W3C
procedure, but this may be acceptable or even desirable, particularly
in the case of a `data' XML document as opposed to a `text'-markup
document.

In \secref{s:problem} below, we review the problem of signing XML, and
the assorted approaches which have, with variously disputed success,
attempted to achieve this.  In \secref{s:procedure}, we describe a
proposed normalization procedure, and the class of documents which is
equivalent under the normalization.  In \secref{s:implementations} we
describe two compatible implementations, for illustration, and briefly
examine performance.  We draw some conclusions
in \secref{s:conclusions}.

\emph{A note on parsers:}
In order to fix terms, if nothing else, we should add here a note on
how XML documents are processed.  The most widely-known XML interface
is probably the `Document Object Model',\footnote{This is managed by
the W3C DOM `Activity' described at \url{http://www.w3.org/DOM/}.}
which represents the entire XML document as a tree in memory, and
defines functions for moving within, and querying, that tree; there is
also a class of so-called `pull' APIs which support passing
deserialized objects to an application.  There is a smaller class of
general-purpose underlying parser models, and the best known of these
is the `push' or event-based parser model best represented by the Java
parser SAX, a \textit{de facto} standard described
at \url{http://www.saxproject.org}.  This parses the token stream
obtained from the serialized XML object and reports its content as a
sequence of `events' corresponding to element beginnings and ends,
textual content, programming instructions, and so on.  Higher-level
APIs such as the DOM will typically use such a parser internally,
which may be swappable by the programmer.  Crucially for
our argument below, there is a rather small set of document events
which such parsers will report.

\section{The problem of normalizing and signing XML}
\label{s:problem}

In a sturdily-reasoned essay, baldly titled `Why XML Security is
Broken', Peter Gutmann has discussed this problem, and suggested that the
approach used by the W3C WG is fundamentally
mistaken~\cite{gutmann04}.  It's a broadly persuasive argument for the
general problem, but in the case of a large category of XML documents,
the problem is not as hard as this analysis suggests, precisely
because we only rarely need to solve the general problem.

The two key points that Gutmann makes are:
\begin{enumerate}
\item All cryptographic signature mechanisms are designed to sign a blob
of bytes.  XML documents are not just blobs of bytes, so there's a
fundamental dislocation here between what's wanted and what's
available.
\item Signature mechanisms are designed to work with \emph{streams} of
bytes, so that it matters from a practical point of view where the
signature is located in a
byte-stream, and that the system knows from the outset which type of
signature it is expected to create or verify.
\end{enumerate}
We do not further discuss point 2 in this paper, on the grounds that
this represents an engineering trade-off between making things easier
for the writer of a signature, or the reader of it.  The mechanism
described in \secref{s:implementations} can cope with this metadata
being located at the beginning of a document or the end.

Gutmann's solution to point~1 is not to normalize at all, but instead to regard the
on-disk or on-the-wire XML document as the blob of bytes to be signed.
This works to some extent, but throws away the mutability of XML, which means that if
one wants to \emph{do} anything with the XML other than
simply admire it, or if one wants to round-trip the XML into and out
of a system which doesn't know about signatures, one is presented
with the dilemma of either abandoning the signature, or else worrying
about how to reproduce exactly the same blob of bytes when the XML is
serialized at some later stage.  One option is to store the original
bytes alongside the parsed form, and make these available for
subsequent inspection; but this redundancy will be at least
inconvenient, probably brittle, and possibly impractical if the data
volume is at all large.

It is part of the point of XML that XML documents are \emph{not} just blobs of
bytes, but represent a structure which is not brittle in the face of
minor textual changes.  This robustness is what enables a rich range of higher-level
applications.  XML processors and editors freely take advantage of
this: it is generally hard to guarantee what flavour of quotes will be
written by an XSLT transformer, or that (formally or practically
insignificant) whitespace will be preserved by XML editors.  The
consequence of this is that even the result of an XML
identity transformation will not match the input byte-for-byte, in a
large fraction of cases; such an identity transformation is what
happens in practice when one opens and immediately resaves a document
in an XML editor, or logs a copy of an input document in an XML
workflow.\footnote{Sect.\,3.4 of
the XSLT standard is as compact and lucid as everything else in that
standard, but it still takes several paragraphs, and implicit
reference to the XPath and XML standards, to describe what text nodes
are and are not removed; the \texttt{xsl:output} element is also
involved in governing the whitespace which appears in XML output.}
Schemas or DTDs can make this mutability
more pronounced, since amongst other things they can license more
extensive syntactic transformations.

An XML document is fundamentally
a tree serialised into bytes; indeed, at the slight risk of being
metaphysical, we might assert that this tree -- in contrast to any
particular serialization into bytes -- \emph{is} the XML document we
wish to sign.  This means that Gutmann's solution, though it is
implementable, works by signing an otherwise insignificant and
transient detail of the XML document.
More practically, this syntactic mutability is
reflected in the fact that applications very rarely operate on
the bytes of a document or stream, but instead on the
\emph{abstracted} content of a document, as exposed via an API such as
SAX, DOM or Expat, or an XSL node-set.  An XML database is free to store an XML
document in any way it likes, as long as it produces an equivalent
document when required.  It is this focus on APIs that shows us how to
give practical force to tbe goal of signing the abstract tree.

\subsection{XML canonicalization}

XML Canonicalization is somewhat complicated -- the process is
summarised in a 14-point list in Sect.\,1.1 of \cite{std:w3cc14n}, which
much of the rest of that document elaborates at length (this is
generally abbreviated `C14N', and appears to be specified independently, and
presumably equivalently, though without mutual cross reference, by
both~\cite{std:w3cc14n} and~\cite{std:rfc3076}; when we refer to
`C14N' below, we are referring specifically to this process).
Much of the detailed
complication, however, arises because the process is still fundamentally
canonicalizing one text file to another text file; it is a
canonicalized \emph{serialization} rather than a canonicalization of the
XML tree itself.  Furthermore, the C14N specification requires that
applications run arbitrary transformation script, which is a potential
and sometimes actual security hole in implementations~\cite{hill07}.

The XML Signature specification (hereafter `xmldsig'; see~\cite{std:xmldsig}
and~\cite{std:rfc3275}) is also a complicated document, servicing an
elaborate set of requirements~\cite{reagle00}.  These requirements
include being able to sign arbitrary fragments of a document, support
for detached signatures, the ability to sign composite documents
listed in a manifest, signatures in the presence of XLink and XPointer
external references, and being able to sign various transformations of
the input including XSLT transformations.  The W3C gathered
a list of significant implementability problems in the standard -- or
rather `topics of interest' -- in a 2007
workshop,\footnote{\url{http://www.w3.org/2007/xmlsec/ws/report.html}}
and these have informed the set of requirements for XML Security
2.0.\footnote{\url{http://www.w3.org/TR/xmlsec-reqs2/}} Whatever
changes emerge, it is clear that the W3C's XML Security standard will
remain a complicated solution to a complicated general problem.

Other pragmatically-motivated XML signing algorithms have been
proposed.  The SAML `SimpleSign' method~\cite{hodges08} (which seems
still to be a draft), and the XML-RSig method (described and discussed
in~\cite{brooke06}; Johannes Ernst's original reference seems to have
disappeared), both address a simplified problem, and both attempt,
like xmldsig, to apply a normalization step to the XML source text --
both of them, that is, are ways of defining the blob of bytes to be
signed.  In doing so, they acquire some robustness against the sort of
accidental transformation which will frequently happen to a document
as it makes its way around the internet.  But this is at best
precarious, and can be frustrated by a step as simple as transcoding
an XML document from ISO-8859 to UTF-8.  These proposals also do not
address the problem of round-tripping a document into and out of an
XML database, or making adjustments to a non-signed part of a document
in an XML editor.

If (we suppose) we cannot simplify the general solution, can we instead
simplify or relocate the problem?

Firstly, it seems likely that, in the majority of cases where XML
applications would benefit from signatures, the requirements are in
fact rather simple, and boil down to not much more than `is this the
same XML document that the sender intended to dispatch?'  In most
cases, little or no transformation or composition of the input
document will be required, nor will, for example, signatures need to
be themselves signable.\footnote{It would be unsurprising if most
current applications of xmldsig do in fact exploit some of these
features; this does not undermine the point: only such demanding
applications are likely to invest the costs of an xmldsig
implementation.}

Secondly, we can relocate the problem by taking seriously the idea
that the object that the signature signs is the parse tree, and not an
XML serialization of it.


The various canonicalization and digital signature implementations
briefly discussed above
use the formal notion of the XML Information Set~\cite{cowan04},
which describes all of the information which a complete XML processor
must preserve and make available to an application (the
`canonicalization' work of the XML Signatures WG is effectively
concerned with defining a single serializaton of this set).

The XML InfoSet is quite elaborate, and includes many of the features
which may be extracted from an XML document.  The key observation for
our present purpose is that there are some features which it
does \emph{not} include: there is no `information item' corresponding
to the ordering of attributes, for example.  The XML C14N view defines
documents as equivalent if they produce the same InfoSet, which means
in effect that they differ only in details which are interchangeable
at or near the lexer level.

The SAX model, however, implicitly defines a much simpler information model for XML,
in terms of just 11 API functions (in the
Java \texttt{org.xml.sax.ContentHandler} interface); the Expat
interface\footnote{\url{http://expat.sourceforge.net}} is structurally
very similar, and supports 12 handler functions (plus handlers for
errors and DTDs); there are exactly analogous APIs in other languages,
which can consequently support the same model.  In both cases, the
simplicity -- compared to the XML C14N view -- arises because the APIs
view of the document is the \emph{application's} view of the document.
An API-level canonicalization therefore implicitly defines a rather
large class of equivalent documents, which blurs lexical details and entity
boundaries, but members of which must, crucially, have the same semantics
downstream.  This canonicalization is insensitive to schemas,
XInclude, xml:base declarations, encodings and other upstream features which
are typically redundant to the parse: if the application sees
something, it's in the canonicalized document; if not, not.

Because this processing is defined at the API level, it is
straightforward to implement it `en passant' during parsing.

Viewed through a SAX or Expat lens, an XML document is a rather simple
thing,\footnote{This is of course a large part of the point of XML.}
which is consequently very simple to normalize, serialize and thus
sign.

\section{A simple-enough normalization procedure}
\label{s:procedure}

Here, we propose a very simple normalization mechanism,
which straightforwardly turns an XML document into a
stream of bytes, in a well-defined, streamable, efficient and partly reversible way.
The resulting stream can be signed straightforwardly by well-understood
tools, and the signature embedded into the XML equally naturally.

The procedure below can be summarised as follows.  Starting with an
arbitrary XML document~$x$, we can obtain from this a binary blob~$b$
via
\[
b = \Norm(\ESIS(x)),
\]
where the $\ESIS$ step is a textual representation of
the \emph{parse-tree} of the document~$x$ (described
in \secref{s:esis} below), and the $\Norm{}$ step consists of a
transformation of that text into a unique binary blob (described
in \secref{s:normalizing}).  Of course these steps can naturally be combined
into a single one in practice.  The blob~$b$ can then be signed,
$s_x=\Sign(b)$, and the result either distributed alongside the
original document~$x$, or included within it in
a \texttt{<?signature...?>} PI.  Section~\ref{s:implementations}
describes the implementation of these steps in~C and Java;
an implementation would be straightforward in any language which
possesses an XML parser, and it might even be possible (as an amusing
exercise if nothing else) to implement it in XSLT if one's tastes ran
that way.

Although the two steps here may be conceptually similar to the C14N
case, there are two important differences.

Firstly, the normalization rules in $\Norm()$ are fewer,
and all but one are nearly trivial; the procedure builds upon the
observation that a good deal of normalization is, in effect, done for
free by a conformant XML parser, in processing whitespace and incorporating
entities, and need not be re-specified.

Secondly, the operations $\Norm$ and $\ESIS$ are naturally defined in
terms of the events produced by a SAX-type parser, and so can be
implemented \emph{during a parse}, as a side-effect of passing the
parse events to the application (this mode is available in the
implementations described in \secref{s:implementations}).  In contrast
the Gutmann approach $s_g=\Sign(x)$ is defined in terms of the bytes
of the unparsed XML document, which makes it at least troublesome to
both validate the document and use it: if one wants to do both, it may
be necessary to scan the document twice.

The simple normalization $\Norm\circ\ESIS$ turns the XML:
\begin{verbatim}
<doc>
<p class='foo'>Hello</p>
  <p> there
chum
</p>
</doc>
\end{verbatim}
into the normalized form:
\begin{verbatim}
(doc
Aclass foo
(p
-Hello
)p
(p
- there chum
)p
)doc
\end{verbatim}
The bytes comprising this normalized form can then be signed, and the signature reinserted
into the original XML, or else made available as the parsed XML is passed
downstream.

The following document has the same normalized form as the previous
one, but includes a PGP signature block which can be used to verify
it.  The signature here is the
ASCII-armoured~$s_x=\Sign(\Norm(\ESIS(x)))$, not Gutmann's~$s_g=\Sign(x)$.
\begin{verbatim}
<doc><p class="foo">Hello</p><p> there
chum          </p>
<?signature armor='-----BEGIN PGP SIGNATURE-----
ABC123....
-----END PGP SIGNATURE-----'?></doc>
\end{verbatim}

\subsection{An extended ESIS format}
\label{s:esis}

The textual representation is inspired by the ESIS format, which
was originally defined, in the 90s, as the output of the \texttt{sgmls}
program~\cite{clark98}.  The
original point of the format was that it should be easy for downstream
tools to parse.  The point here is that it turns an XML file into an
unambiguous sequence of lines and, further, that this can then be turned
into an unambiguous byte-stream by a simple normalization operation.

There isn't a complete overlap between the ESIS and the SAX/Expat
model.  All the differences here are extensions rather than changes.

The output consists of a sequence of lines.  Each line consists of a
start character indicating which type of output record it represents,
followed by one or more arguments.  There are always the same number
of arguments, separated by a single space.  The extended syntax is
described in \tableref{t:esis2}.  Each start element event
is \emph{preceded} by the set of attributes on that event.
\def\meta#1{$\langle$\textit{#1}$\rangle$}
\begin{table}
\begingroup
\def\extn{${}^*$}
\let~\textvisiblespace
\begin{tabular}{ll}
\texttt{M}\meta{prefix}~\meta{uri}&start prefix mapping\extn\\
\texttt{m}\meta{prefix}&end prefix mapping\extn\\
\texttt{A}\meta{attname}\texttt{~CDATA~}\meta{attvalue}
        &declare attribute\\
\texttt{B}\meta{namespace}~\meta{localname}~\texttt{CDATA}~\meta{attvalue}
        &declare namespaced attribute\extn\\
\texttt{(}\meta{name}&start element\\
\texttt{[}\meta{namespace}~\meta{localname}&start namespaced element\extn\\
\texttt{)}\meta{name}&end element\\
\texttt{]}\meta{namespace}~\meta{localname}&end namespaced element\extn\\
\texttt{-}\meta{text}&character content\\
\texttt{=}\meta{text}&ignorable whitespace\extn\\
\texttt{?}\meta{target}~\meta{data}&processing instruction\\
\texttt{X}\meta{name}&skipped entity\extn\\
\end{tabular}
\endgroup
\caption{\label{t:esis2}Extended ESIS output.  The extensions to the
  original ESIS output are marked with a star.}
\end{table}

Notes to \tableref{t:esis2}:
\begin{description}
\item[M and m] The \meta{prefix} here is the XML prefix which is mapped to
  the given namespace URI.  In the case where the default prefix is
  mapped, the prefix is the empty string.
\item[string values]In the \meta{attvalue}, \meta{text} and PI \meta{data}
  fields, any line-end characters are escaped as
  \verb-\n-, \verb-\r-, \verb-\u0085- or \verb-\u2028- as appropriate.
  In the case of \meta{attvalue}, the string will have been normalized
  by the XML parser, as described in Sect.~3.3.3 of the XML 1.0 and 1.1
  specifications.  Note also that, as a result of the line-end
  normalization rules (XML specs, Sect.~2.11), line-end characters
  other than U+000a can appear only as the result of the expansion
  of a character reference.  This step is handled upstream by the XML
  parser and so does not need to be specified here.
\item[Ignorable whitespace]This record can be generated by the
  presence of whitespace in places where a DTD does not allow mixed
  content.
\item[Processing instructions] The \meta{data} value is the content of
  the processing instruction after removing any whitespace which
  follows the PI target, and removing any whitespace which precedes
  the PI end marker (\verb-?>-).
\end{description}

\subsection{Normalizing the ESIS output}
\label{s:normalizing}

Given a representation of XML in this textual form, we normalize it using
the following procedure:
\begin{enumerate}
\item Ignorable whitespace (`\texttt{=}'), skipped entities (`\texttt
  X'), and start and end prefix mappings (`\texttt M' and `\texttt m')
  are discarded.
\item All of the output is encoded to bytes as UTF-8.
\item Each of the lines is terminated by a CR LF pair (ie, bytes \texttt{0xd 0xa}).
\item Attribute records (`\texttt A' and `\texttt B') are ordered, as
  byte-strings, on output (this implies that all of the `A' records
  appear before the `B' records).  Each of the `\texttt B`,
  `\texttt{[}' and `\texttt{]}' records include a \meta{namespace}
  URI; these should be unchanged from the form in which they appear in
  the XML document.  This is consistent with the stipulation
  of \cite[\S2.3]{std:xml-names} (and cf.\ \cite[\S6.2]{std:rfc3986})
  that two namespace URIs `are identical if and only if the strings
  are identical, that is, if they are the same sequence of
  characters.'
\item Attributes which are in the
  namespaces \url{http://www.w3.org/XML/1998/namespace}
  or \url{http://www.w3.org/2000/xmlns/} -- that is, those attributes
  with a \texttt{xml:} or \texttt{xmlns:} prefix -- are discarded
  (APIs typically do not report these as element attributes).
\item All of the attribute records are listed as \texttt{CDATA},
  irrespective of any type declared in a DTD.
\item Successive `\texttt-\meta{text}' events are merged
  before the following step.
\item The \meta{attvalue}, \meta{text} and PI \meta{data} values are
  normalized by collapsing all runs of one or more whitespace
  characters to a single space (U+0020).  This happens irrespective of whether
  the whitespace character was present in the input XML as a character
  or as a character reference.  This step is very similar to the
  `attribute-value normalization' of non-CDATA attributes as described
  in Sect.~3.3.3 of the XML specifications, but without the exceptions
  for whitespace character references.
  The `whitespace' characters are:
  \emph{all} of the characters below U+0020,
  plus the NEL character (U+0085)
  and the Unicode line-separator character (U+2028), and no others
  (the characters below U+0020 other than tab, newline and
  carriage-return are either illegal characters, in XML 1.0, or
  `discouraged', in XML 1.1, and so should never appear in the input,
  but are included here for completeness).
  The characters U+0085 and U+2028 are special in XML~1.1 but not in~1.0.
  The other Unicode whitespace characters (category `Zs')
  are not taken to be `whitespace' in this sense.
\item If a \meta{text} record is empty after this normalization, or
  contains only whitespace, it is discarded.
\item Any processing instruction which has a \meta{target} of
  \texttt{signature} is removed.
\end{enumerate}

The result of this is to transform and normalize XML as illustrated in \figref{f:transform}.

\begin{figure}
\begin{tabular}{lll}
XML&Un-normalized&Normalized\\
\hline
\begin{minipage}{0.33\textwidth}
\begin{verbatim}
<doc><pfx:p class='foo'
  xmlns:pfx="urn:NS"
  pfx:att='bar'
>Hello</pfx:p>

<p>   &amp;&#xD;goodbye,
chum</p>
  </doc>
\end{verbatim}
\end{minipage}
&
\begin{minipage}{0.33\textwidth}
\begin{verbatim}
(doc
Mpfx urn:NS
Aclass CDATA foo
Burn:NS att CDATA bar
[urn:NS p
-Hello
]urn:NS p
mpfx
-\n\n
(p
-   &\rgoodbye,\nchum
)p
-\n
)doc
\end{verbatim}
\end{minipage}
&
\begin{minipage}{0.33\textwidth}
\begin{verbatim}
(doc
Aclass CDATA foo
Burn:NS att CDATA bar
[urn:NS p
-Hello
]urn:NS p
(p
- & goodbye, chum
)p
)doc
\end{verbatim}
\end{minipage}
\end{tabular}
\caption{\label{f:transform}The transformation of XML into pseudo-ESIS
form, and its subsequent normalization.}
\end{figure}


In the normalized form, the prefix mappings have been removed
(the prefixes are not semantically important),
whitespace has been collapsed within the `\texttt -\meta{text}' lines,
and two all-whitespace `\texttt -\meta{text}' records have been
removed.

A programming instruction (PI) with \meta{target} `signature' (that is a PI
of the form \texttt{<?signature ... ?>}) is handled specially.
The \meta{data} portion of this PI consists of a sequence of
key-value pairs, where each value is enclosed in single or double quotes.  The
only keys defined so far are:
\begin{description}
\item[\texttt{algorithm}]
This indicates the type of signature.  Possible values
include \texttt{pgp} to indicate PGP signatures,
and \texttt{md5}, \texttt{sha1} or similar to indicate cryptographic hashes.
\item[\texttt{content}]
This indicates the cryptographic hash of the content, or (depending on
the \texttt{algorithm} chosen) the PGP-armoured output of a PGP/GPG signature
(that is, starting with \texttt{-----BEGIN PGP SIGNATURE-----}).
\item[\texttt{target}]
This indicates the element which is to be signed. It may have one of
the values \texttt{/} or \texttt{following::*[1]}, indicating
respectively the whole document or the XML element immediately
following the signature PI.  In the absence of this attribute, the
signature is taken to refer to the whole document.  The permitted
attribute values are these literal strings.  They are indeed
syntactically XPath specifiers, but there is no implication that an
arbitrary XPath may be provided here.
\end{description}
The signature in question is a signature of the entire normalized
output, $\Norm(\ESIS(\text{\emph{$\langle$target$\rangle$}}))$, taken as a blob of bytes.

The signature PI may appear at any point in the input XML, and
identifies a signature for the complete XML document which includes
it.  In particular, it does not matter in this scheme whether this PI,
indicating the algorithm and signature, is at the beginning of the
document or at the end; having it at the beginning is slightly harder
to generate, but much easier to verify afterwards.  We have not
defined them here, but it is easy to see how modest developments of this
scheme might use the \texttt{<?signature...?>} PI to indicate either
a different element to sign, or a different algorithm to use, or might
use multiple PIs to separate the algorithm parameters at the beginning
of an element from the signature result at the end.

\subsection{The equivalence class of documents under this normalization}

This is a rather aggressive normalization: it defines a class of
equivalent XML documents which is somewhat larger than the equivalence
class implied by the C14N procedure.  In particular, (i)~documents which
differ only in whitespace are deemed equivalent, and more radically
(ii)~all details of internal and external entities, and their provenance,
are lost.  The rationale for this is that the distinctions represented
by~(i) are semantically insignificant in a very large variety of
important cases, and those represented by~(ii) are in any case
invisible to a SAX/Expat client application.  This does mean that
documents \texttt{<d><p>a</p>  <p>b</p></d>}
and \texttt{<d><p>a</p><p>b</p></d>} (the first has a space between
the `p' elements) would be deemed equivalent; we
assert that this is unlikely to be a problem in practice.

This means that, although the transformation here is invertible, in
the sense that it can be reversed to produce an XML document which is
equivalent to the original in the sense described, that reconstructed
document may look, at first glance, rather different from the
original.

By making these equivalences, we avoid a very large fraction of the
complications of the C14N algorithm, and produce a blob of bytes which
is a natural object to receive a digital signature.

\section{Implementations}
\label{s:implementations}

The process described here has been illustratively implemented in a
Java class library and a C library, which are available at
\url{https://bitbucket.org/nxg/xoxa}.  This article corresponds to
version 0.3.1 of the implementation.

Both libraries implement the transformation and normalization steps
described here, as command-line applications as well as an API.

The Java library provides (amongst other classes) a
class \texttt{SigningXMLReader} which subclasses the
SAX \texttt{XMLFilterImpl} class, and so implements the
SAX \texttt{XMLReader} interface, and which in addition generates and
checks GPG signatures within the input XML.  It can thus be swapped in to an
application in place of such a SAX reader, and work as that reader
does, with the exception that, after the reader has completed parsing
the input document, it can be queried for details of any signature
found within the source XML, within a \texttt{<?signature ...?>} PI,
including the verification status of the signature.

The C library similarly provides an API for obtaining the normalized
and unnormalized versions of an XML input file, as well as an
interface for obtaining one or other cryptographic digest of the
normalized form.  In addition, and analogously to the Java library, it
provides an API which exactly mirrors the
Expat \texttt{XML\textunderscore *} functions, in the sense that for
each Expat function such as \texttt{void XML\textunderscore
SetStartElementHandler (XML\textunderscore Parser, ...)}, there is a
function named \texttt{void Xoxa\textunderscore SetStartElementHandler
(Xoxa\textunderscore Parser, ...)}  with the same effect, which can
therefore be dropped in as a replacement.  The difference is that it
is then possible to query the \texttt{Xoxa\textunderscore Parser}
object to obtain a cryptographic digest of the normalized input XML
(the C implementation supports cryptographic digests but not, so far,
PGP signatures).

\begin{table}
\begin{centering}
\begin{tabular}{r|ccc|c}
\textbf{size/MB}&\textbf{time}(expat)/s&\textbf{time}(xoxa)/s&\textbf{ratio}&\textbf{time}(digest)/s\\
\hline
11      & 0.370 & 0.781 & 211\% & 0.0437\\
35      & 1.054 & 2.28 & 216\% & 0.122 \\
117     & 3.51  & 7.57  & 216\% & 0.395 \\
351     & 10.65 & 23.0  & 216\% & 1.247 \\
1172    & 35.8  & 76.2  & 213\% & 4.03
\end{tabular}\\
\end{centering}
\caption{\label{t:perf-c}Performance on identity transform, with Xoxa-C.
The columns show the sizes of the input files, the times to do an identity
transform using the native Expat interface, and with the replacement
Xoxa interface, including a digest calculation.  The last column is
the time required to perform just a digest calculation on the
Xoxa-normalized file (that is \texttt{openssl sha1 file.norm}, after
first generating the file with \texttt{xoxa file.xml >file.norm}), with no
significant output.  All of these figures are averaged over three
instantiations of the random test input.}
\end{table}

In \tableref{t:perf-c} we illustrate the relative performance of an
identity XML transform (that is, parsing XML, and then immediately
serializing and writing it) using alternatively the C-based Expat
interface (that is, the \texttt{XML\textunderscore *} functions) and
the C-based Xoxa one (that is, replacing
the \texttt{XML\textunderscore *} functions by the
corresponding \texttt{Xoxa\textunderscore *} ones).  In each
case, the programs are processing randomly generated
XML.\footnote{Generated by XMark,
see \url{http://www.xml-benchmark.org}.  This produces
non-pathological XML with a roughly even mixture of markup and text
content, and with random nesting depths up to 10 levels deep.}  There
is little variance in the timings, and the
increase in processing time in each case is very nearly linear with
the size of the input XML file.  The extra layer of
indirection at parse time, plus the calculation of the SHA1
cryptographic digest, roughly doubles the run time in this simple
case.  This is a worst case, however: a real application would
do significantly more with the parsed XML than simply write it out, so
that we can see that the cost of the extra processing would be minor in a
real application.  The processing in each case is CPU-bound.

\begin{table}
\begin{centering}
\begin{tabular}{r|ccc|c}
\textbf{size/MB}&\textbf{time}(default)/s&\textbf{time}(xoxa)/s&\textbf{ratio}&\textbf{time}(gpg)/s\\
\hline
11      & 0.713 & 1.13  & 159\% & 0.0843 \\
35      & 1.092 & 2.19  & 200\% & 0.237  \\
117     & 2.43  & 5.75  & 237\% & 0.835  \\
351     & 6.10  & 15.79 & 259\% & 2.48   \\
1172    & 19.24 & 50.0  & 260\% & 8.39
\end{tabular}\\
\end{centering}
\caption{\label{t:perf-j}Performance on identity transform, with
Xoxa-J; the columns are as in \tableref{t:perf-j}, except that the
final column is the time taken to verify GPG signatures of a normalized
input file (\texttt{gpg --verify file.sig file.norm}).}
\end{table}

In \tableref{t:perf-j} we illustrate the performance of an equivalent
task using the Java implementation.  In this case the identity
transformation is implemented by
\begin{verbatim}
Result res = new StreamResult(System.out);
Transformer t = TransformerFactory.newInstance().newTransformer();
t.transform(src, res);
\end{verbatim}
(the Java classes are from the \texttt{javax.xml.transform} package or
its subpackages).  Here, the \texttt{src} is created in the default
case by
\begin{verbatim}
Source src = new SAXSource(new InputSource(...));
\end{verbatim}
and in the Xoxa case by
\begin{verbatim}
XMLReader rdr = uk.me.nxg.xoxa.SigningXMLReader.getXMLReader();
Source src = new SAXSource(rdr, new InputSource(...));
\end{verbatim}
with no other differences.  The \texttt{rdr} object can then be queried, after
the parse, to obtain information about the signatures.
As with \tableref{t:perf-c}, the GPG run times are linear with the input
source size, but the Java run times only become linear above 100MB.
As with the C case, the run time in the Xoxa case is increased by a
factor of two or three compared with the default case; as before, this
would become less significant in a real application which was doing
something more substantial with the input.

\section{Conclusions}
\label{s:conclusions}

Cryptographically signing XML, and normalizing it prior to signing,
are forbiddingly intricate problems in the general case.  This is
largely because of the complexities of the XML Information Set, and
because of the decision to define the normalization in terms of the
input XML syntax rather than the parsed structures which are more
integral to the document.

In this article, we have described a transformation from an XML
document to a readily-signed blob of bytes which is:
\begin{enumerate}
\item straightforwardly described (partly because it sits on top of
well-defined existing processes such as XML parsing);
\item reasonably straightforward to implement; and
\item robust against trivial but likely changes to the source file,
which might occur during transmission or storage.
\end{enumerate}
The procedure is naturally portable, and has been illustratively
implemented in both Java and~C.


\end{document}